\documentclass[pre,twocolumn,showpacs]{revtex4}
\usepackage{graphicx}
\usepackage{amsmath}
\usepackage{amsfonts}
\usepackage{psfrag}
\newcommand{\kbar}{\mathchar'26\mkern-9muk}
\begin{document}
\title{Quantum and Classical Effects in the Two-Frequency Kicked Rotor with Variable Initial Phase}
\author{T. G. Mullins}
\author{A. J. Hilliard}
\author{M. P. Sadgrove}
\author{M. D. Hoogerland}
\author{A. S. Parkins}
\author{R. Leonhardt}
\affiliation{Department of Physics, The University of Auckland, P.O. Box 92019,
Auckland}
\begin{abstract}
We present an investigation into effects exhibited by the two-frequency kicked rotor.  Experiments were performed and in addition quantum and classical dynamics were simulated and compared with the experimental results.  The experiments involved pulsing the optical standing wave with two pulsing periods of differing frequencies and variable initial phase offset.  The ratio of pulsing periods was sampled for rational and irrational values for different experimental runs.  In this paper we present these results and examine the measured momentum distributions for the cause of any structures that are seen in the energy as the initial phase offset is changed.  Irrational ratios exhibit no significant quantum effects, whereas rational ratios show dynamical localisation (DL) for certain values of the initial phase.  However, most of the observed structure is found to be due to classical effects, in particular KAM boundaries, and is therefore not of uniquely quantum origin.
\end{abstract}

\maketitle

\section{Introduction}

One of the signatures of quantum behaviour in the atom optics kicked rotor (AOKR) is dynamical localisation (DL).  For a classically chaotic phase space, the energy of the classical kicked rotor increases approximately linearly with time and grows indefinitely.  However, in the corresponding quantum system, the initial energy diffusion is approximately linear, but  after some characteristic (localisation) time the energy growth is curtailed \cite{Casati,Moore}.  This halting of energy growth is due to quantum mechanical interference and is expected to be very sensitive to changes in periodicity, as the quantum correlations are sensitively dependent on the time between successive kicks \cite{Siri2001}.

It has been shown experimentally that DL is sensitive to decoherence in the form of spontaneous emission \cite{NelsonPRL,Raizen1998a}.  Previous work has also shown that in the presence of a small amount of period noise, DL is no longer observed \cite{RaizenPeriod,Mark,andypandy}.  Experimental results in \cite{Frenchgroup}, where two-frequency driving was employed, suggest that DL may only be observed when the ratio of the driving frequencies is rational.  These same results also exhibit sub-Fourier resonances \cite{SubFourier}.

In the experiment performed in \cite{Frenchgroup} the ratio, $r'$, of the pulsing frequencies of the two pulse-trains was changed and the number of zero velocity atoms measured.  The time between the first pulse of each pulse-train, relative to the period of the second pulse-train (the phase), was set to $52^{\circ}$ (where $360^{\circ}$ corresponds to one period of the second pulse train) to avoid pulse overlaps.  At highly rational values of $r'$ ($1/2$, $1$, $2$, etc.) a large increase was observed in the number of zero velocity atoms.  This was attributed to DL, as the same effect was not observed in simulations of the classical $\delta$-kicked system.

If pulse overlaps were able to be taken into account, the experiment would not be limited to any particular initial phase and the effective periodicity could be altered by changing that phase.  In particular, investigating phase dependence would ensure that the results in \cite{Frenchgroup} are not merely due to a particular choice of parameters, and that other effects do not come into play.

In this paper we again consider the case where the driving may be periodic or quasi-periodic.  We add a second pulsing frequency (period $T_2$) to our primary pulsing frequency (period $T_1$), as in \cite{Frenchgroup}.  Since DL is extremely sensitive to changes in periodicity, very small changes in the initial phase offset, $\psi_0$, should affect the amount of DL observable on the finite experimental timescale.  This effect should occur for commensurate frequencies (as DL occurs for pulse-trains with sufficient periodicity), but not for incommensurate frequencies.

The ratio $r=T_2/T_1$ defines the period of the second pulse-train.  The rational or irrational nature of $r$ should determine whether or not DL is observable on a particular time scale.  The ``more'' irrational the value of $r$, the longer the onset of DL is expected to take.  In practice it is only possible to have rational values of $r$ and thus irrational ratios may only be approximated by a close, rational ratio \cite{Kim}.  This approximation should still give similar results to true irrational ratios on the timescale investigated in this paper: for a given $r=a/b$, which is an irreducible fraction, the periodicity of the system cannot be seen before at least $b$ kicks of the primary pulse-train.  Thus if the number of kicks from the primary pulse-train is not large enough, compared with $b$, then DL cannot be observed and $r$ may be considered to be irrational on that timescale.

We investigate the effect that changing the initial phase offset has on DL.  We also present evidence of classical effects, which in fact dominate the dynamics and are also strongly affected by changing the initial phase offset.

\section{Atom optics kicked rotor with two-frequency driving}
\label{sec:AOKR}

The Hamiltonian for an AOKR kicked with two driving frequencies, corresponding to periods $T_1$ and $T_2=rT_1$, is given, in scaled units, by:
\begin{eqnarray}
\label{eq:ham1} \hat{H} = \frac{\hat{\rho}^2}{2} & + & \cos(\hat{\phi}) \left[\kappa_1 \sum_{n=0}^{N} k_1(\tau-n) \right. \nonumber\\
& + & \left. \kappa_2 \sum_{m=0}^{M} k_2 \left( \tau-(rm + \alpha_0) \right) \right],
\end{eqnarray}
where $\hat{\phi}=2 k_L \hat{x}$ and $\hat{\rho}=(\kbar / 2 \hbar k_L) \hat{p}$ are operators representing the (scaled) atomic position and momentum respectively, $\kappa_1$ ($\kappa_2$) is the kicking strength of pulse-train 1 (2) (i.e. the area under the pulse),  $k_1$ ($k_2$) is the pulse shape function for pulse-train 1 (2), $\alpha_0$ is the scaled time between the first pulse of pulse-train $1$ and the first pulse of pulse-train $2$ ($\psi_0=\alpha_0 \cdot 360^{\circ}$ defines the initial phase offset) and $\tau=t/T_1$ is the scaled time. We note the scaled commutator relationship $[\hat{\phi}, \hat{\rho}] = i\kbar$, where $\kbar=4 \hbar k_L^2 T_1 / m$ is the effective Planck's constant.  Note that $1 \hbar k_L \equiv \kbar/2$ in scaled units.  We also define $M+N=N_{\rm{tot}}$ as the total number of kicks.  Note that our definition of the ratio is different to that in \cite{Frenchgroup}, where $r'=f_2/f_1 \equiv T_1/T_2=1/r$.

$N_{\rm{tot}}$ may be chosen and $M$ and $N$ adjusted such that the temporal length of each individual pulse-train is minimised.  As pulse overlaps may occur whilst changing the phase and the ratio, $N_{\rm{tot}}$ may be different from the number of individual pulses obtained from the sum of the two pulse-trains, $N_{\rm{res}}$.

\section{Experimental setup}
\label{sec:ExpSet}

We realize the AOKR by first trapping and cooling approximately $10^5$ Caesium atoms to around $5 \mu$K in a MOT.  The atoms are then subjected to a pulsed standing wave of laser light (the kicking beam).  The kicking beam is obtained by using a $150$mW (slave) diode laser that is injection locked to a frequency stabilised (master) diode laser.  The frequency of the master laser may be adjusted over a range of about $\pm 2$ GHz with respect to the D2 line at $852$ nm.  The output from the slave laser passes through an acousto-optic modulator (AOM) that is used to produce optical pulses, and this light is coupled into an optical fibre and transported to the MOT.  The kicking laser has a maximum available power of $50$mW just before the MOT.  The standing wave is created by retroreflecting the light and has an intensity distribution that is well described by a Gaussian with width 0.72mm at the position of the cloud.

An experiment consists of extinguishing the MOT beams and quadrupole magnetic field, exposing the atoms to a resultant pulse train and allowing the cloud to expand for a fixed time ($12$ms beginning from the moment the MOT beams are turned off) so as to increase the spatial resolution and to allow for time of flight measurements.  Finally an optical molasses is applied to the atoms, ``freezing'' them in place and causing them to fluoresce, and an image of the spatial distribution of this fluorescence signal is recorded using a CCD camera.  From this atomic position distribution, and with knowledge of the time of flight and the initial position distribution, the momentum distribution, and hence the energy of the cloud, may be deduced.

For two-frequency pulse-trains, it is possible for the optical pulses to overlap and as such it is important that we quantify the exact nature of the pulse train.  The switching of the AOM is controlled by a programmable pulse generator (PPG).  The PPG has a timing resolution of $40$ns and can store a pulse-train of up to $2.6$ ms in length.  Pulse trains of arbitrary shape may be uploaded to the PPG by a computer.  If an overlap was to occur, the maximum laser intensity at the MOT would need to be twice as large as for the non-overlapping case.  For this reason the maximum laser power used was $25$mW for non-overlapping pulses.

The pulses output from the PPG were square, with negligible rise and fall time.  The period of the first pulse-train was chosen to be $30 \mu$s ($\kbar = 3.12$) and the pulse-width for both trains was set to $480$ns.  For all experiments the height of each pulse-train was chosen to be identical, and thus $\kappa_1=\kappa_2$ (although overlaps may cause a partial doubling of the height and/or lengthening of a resultant pulse).  The number of pulses was chosen such that $N_{\rm{tot}}=30$.  The shape of the pulses at the MOT is discussed in section \ref{sec:CSims}.

In order to reduce the uncertainty due to small fluctuations in the experimental setup (kicking laser power, vibrations, slight differences in the initial cloud, etc.), three images were taken for each set of parameters.  After an experimental run was completed, it was repeated and an average was taken over all sets of results, giving a measure of the repeatability of the experiment.  As a final test of repeatability, one set of experiments was entirely repeated two weeks later.  The results from this second set of experiments were identical to the results from the first within the experimental error.

\section{Simulations}
\label{sec:Sims}

Both classical and quantum numerical simulations were performed for this system, based on the Hamiltonian given by Eq. (\ref{eq:ham1}).  The Jacobi elliptic function solution for a classical rotator was used to simulate the classical system and the quantum system was simulated using the Monte Carlo Wavefunction method.  In both cases, several non-ideal aspects of the physical experiment were modelled.  These factors were spontaneous emission, the spread in kicking strength due to the magnetic sublevels of Caesium and the finite sizes of the kicking laser and the cloud, and the shape of the optical pulses.

\subsection{Classical Simulations}
\label{sec:CSims}

For the classical simulations, free evolution was modelled by the map:
\begin{eqnarray}
\phi_{\rm{[n+1]_{-}}} & = & \phi_{\rm{n_{+}}} + (\tau_{\rm{[n+1]_{-}}} - \tau_{\rm{n_{+}}} )\rho_{\rm{n_{+}}},\\
\label{eq:classfreemap}
\rho_{\rm{[n+1]_{-}}} & = & \rho_{\rm{n_{+}}},
\label{eq:classfreemap1}
\end{eqnarray}
where $\tau_{\rm{[n+1]_{-}}} - \tau_{\rm{n_{+}}}$ refers to the scaled time between the end of the $n^{th}$ pulse and the start of the $(n+1)^{th}$ pulse.  During the time that the potential was on, the Jacobi elliptic function solution for the classical rotator was evaluated numerically.

Both the initial position within one potential well and the initial momentum of the particle are chosen at random for each trajectory.  The position is chosen from a uniform distribution in the range $[-\pi,\pi)$ and the momentum is chosen from a Gaussian distribution with a standard deviation equal to that given by our measurement of the temperature.

The finite size of our cloud, as well as our kicking beam, have also been taken into account in the simulation.  This is done by randomly selecting a position from a Gaussian distribution that has the same $\sigma$ as our cloud and evaluating the intensity at that position.  Knowing the relative sizes of the beam and the cloud, it is possible to scale the kicking strength accordingly.

Spontaneous emission is taken account of in the simulation in the following manner:  A number is randomly chosen from a uniform distribution between $0$ and $1$.  If that number is below our calculated spontaneous emission rate per pulse, then a spontaneous emission event is deemed to occur during the pulse.  The change in momentum due to the spontaneous emission is chosen from a uniform distribution in the range $[-\hbar k_L,\hbar k_L)$ (corresponding to $[-\kbar/2,\kbar/2)$ in scaled units).

It is important that enough timesteps are taken throughout a pulse to accurately sample the pulse shape.  It is also important in the case of overlapping pulses that more time steps are taken as the overlaps may cause a lengthening of a particular resultant pulse.

The rise and fall time of the pulses used in experiment was also taken account of in simulation.  The pulse shape was measured just before the MOT, using a photodiode with a $1$ns response time. As in reference \cite{Steck2000}, the pulses are described by the formula:
\begin{equation}
k(t) = \frac{k_{max}}{2} \left[ \rm{erf} \left( \frac{(t-t_1) \sqrt{\pi}}{\delta t_1} \right) - \rm{erf} \left( \frac{(t-t_2) \sqrt{\pi}}{\delta t_2} \right) \right],
\label{eq:pulseshape}
\end{equation}
where $k_{\rm{max}}$ is the maximum height of the pulse, $t_2 - t_1$ is the full width at half maximum, $\delta t_1$ is the rise time (defined such that a straight line going from $0$ to $100\%$ of the pulse height in time $\delta t_1$ matches the slope of the rising edge at the half-maximum point), $\delta t_2$ is the fall time (defined similarly to the rise time) and $\rm{erf} (x)$ is the error function.  It should be emphasised that, as in \cite{Steck2000}, this is merely an empirical description of our pulse shape and does not arise from physical considerations.

When the height of the pulse rises to around $10\%$ of its maximum, the pulse is deemed to have begun.  Similarly when the height of the pulse has fallen to less than about $10\%$, the pulse is deemed to have ended.  Everywhere else the pulse height is set to $0$ and free evolution is assumed.

The parameters obtained for our pulses were: $\delta t_1=104$ns, $\delta t_2=121$ns and $t_2-t_1=396$ns.  These result from the measured power values just before the MOT with the PPG outputting a square pulse of duration $480$ns.

As a trade-off between sampling accuracy and CPU time, it was decided that at least $16$ time steps would occur during each non-overlapping pulse.  If an overlap occurs so that a pulse is actually longer, then more timesteps are chosen such that the length of the timesteps are as close as possible to those for the non-overlapping case, but are equal or shorter in stepsize.  We found, from performing simulations, that the energy as a function of the number of steps (for a given number of kicks) varied only a small amount when the number of time steps was $10$ or greater.

Implementing this pulse shape, we found that the energy output from the simulation was lower compared with the result using a square pulse shape.  This lower energy was in better agreement with the experimentally obtained energies than the energies obtained using a square pulse in the simulation.  Also the structures obtained (if there were any present to begin with) were slightly smoothed out and peaks and dips were broadened (again only by a small amount) in comparison with using a square pulse of the same area.  

To ensure the area under the real pulse was the same as the square pulse, the maximum height of the pulse (denoted as $k_{\rm{max}}$ in Eq. (\ref{eq:pulseshape})) was adjusted appropriately.  For a square pulse the kicking strength is given by $\kappa=k \alpha$, where $\alpha$ is the temporal width of the pulse and $k \equiv k(\tau)$ is the height of the pulse.  When implementing the real pulse shape, $\kappa$ still has the same physical significance, but must now be expressed in the more general form:
\begin{equation}
\kappa = \int_{-\infty}^{\infty} k(\tau) d\tau,
\label{eq:kappareal}
\end{equation}
which for $k(\tau) = \rm{const}.$ reduces to the form given previously for the square pulse.

Due to the fact that we are dealing with a classically chaotic process, slight differences in initial conditions give wildly different final energies.  As such, a very large number of trajectories need to be calculated so as to accurately map out the phase space.  In the classical simulations $10,000$ or $40,000$ trajectories were evolved (more trajectories were needed for larger kicking strengths to give similar errorbars to the lower kicking strength case).  This was found to be an acceptable trade-off between simulation time and accuracy.

\subsection{Quantum Simulations}
\label{sec:QSim}

For the quantum simulations the method used was similar to that of ref. \cite{Daley2001}.  A momentum is chosen from a Gaussian distribution in the same way as in the classical simulation.  This momentum is then rounded to the nearest momentum ladder state and this is the initial state that is evolved.  The difference between the momentum of the ladder state and that of the chosen initial momentum is stored as the quasi-momentum $q$.

The initial state is evolved according to the Hamiltonian:
\begin{equation}
\hat{H} = \frac{(\hat{\rho}+q)^2}{2},
\label{eq:freeham}
\end{equation}
\noindent in between pulses and:
\begin{eqnarray}
\hat{H} &=& \frac{(\hat{\rho}+q)^2}{2} + k(\tau) \cos(\hat{\phi}) \nonumber \\
& - & i \kbar k(\tau) \frac{\eta(\tau)}{2} \left( 1+\cos(\hat{\phi}) \right),
\label{eq:kickham}
\end{eqnarray}
where $\eta$ is the probably of a spontaneous emission event occuring during one pulse, while the potential is on.  Note that both $k$ and $\eta$ are functions of time (as the intensity of the potential changes as a function of time due to the pulse shape described by Eq. (\ref{eq:pulseshape})). The last term in Eq. (\ref{eq:kickham}) exponentially decreases the norm of the wavefunction so that spontaneous emission may be included in the simulation.  When the norm of the wavefunction drops below a randomly chosen value, a spontaneous emission event occurs during that pulse.

This process is repeated for multiple initial states (typically in the order of $1000$) and an incoherent average is taken over all of these trajectories to obtain the final momentum distribution.

\section{Experimental and simulation results}
\label{sec:ExpRes}

In this section the experimental and simulated results will be discussed.  These have been divided into three sections.  First commensurate frequencies in section \ref{sec:comfreq}, then incommensurate frequencies in section \ref{sec:incomfreq} and finally the energy as a function of the ratio, $r$, in section \ref{sec:ratio}.

\subsection{Commensurate Frequencies}
\label{sec:comfreq}

For rational $r$, giving a periodic pulse-train (which contains a sufficiently large number of kicks for the periodicity to be resolved), coherent quantum effects, such as DL, should be observable.

\subsubsection{$r=1$}

\begin{figure}[!h]
\centering
\includegraphics[height=5cm,width=8cm]{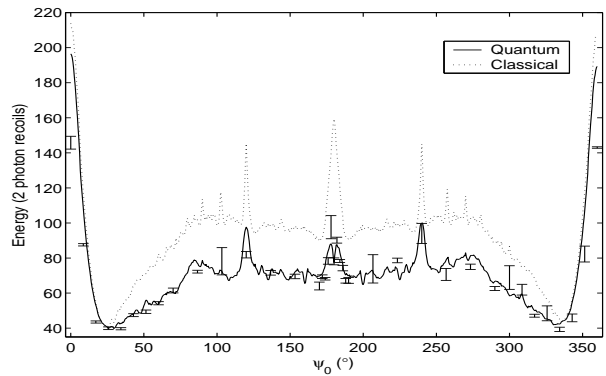}
\caption[Results for $r=1$ and $\kappa=17.7$]{Experimental energies along with quantum and classical simulations for $r=1$, $\kappa=17.7$, $N_{\rm{tot}}=30$ and $\eta=3\%$ as a function of $\psi_0$.  Errorbars are typically $\pm 1.3$ energy units for the quantum simulation and $\pm 1.5$ energy units for the classical simulation.}
\label{fig:highr130}
\end{figure}

Figure \ref{fig:highr130} shows experimental results, along with quantum and classical simulations, for $r=1$, $\kappa_1=\kappa_2=17.7$ and $N_{\rm{tot}}=30$ as the initial phase offset $\psi_0$ is varied.

The experimental and simulation results show that changing the initial phase offset does have a large effect on the resultant energy.  A lot of large scale structure, as well as various resonance peaks, is seen in the experimental results and also in the simulations.

The agreement between the quantum simulation and the experimental results is very good.  The energies from the classical simulations are larger than those of the quantum simulations for most values of $\psi_0$, suggesting the presence of DL.

\begin{figure}[!h]
\centering
\includegraphics[height=5.5cm]{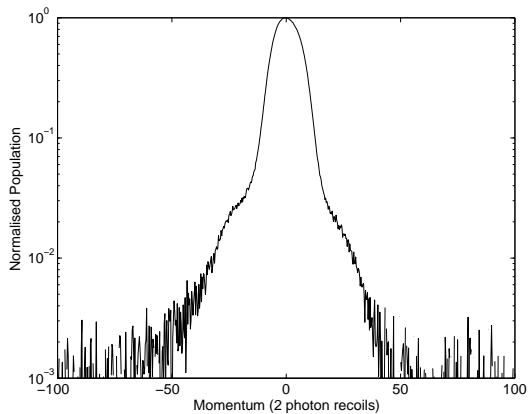}
\caption[Typical KAM momentum distributions for $r=1$]{Experimental KAM boundary momentum distribution for $r=1$, $\psi_0=34.3^{\circ}$, $\kappa=10.1$ and $N_{\rm{tot}}=30$.  The shoulders are a symptom of KAM boundaries and appear at the momentum values that KAM boundaries occupy. Trajectories initially inside the boundary (essentially all trajectories are inside for a cloud temperature of around $5\mu$K) have difficulty diffusing through it, hence the fairly sharp drop off in population.  Note that this distribution has not been deconvolved with the initial position distribution, but this has only a slight effect on its shape.}
\label{fig:lowr130KAM}
\end{figure}

Most of the large scale structure may be explained by the presence of KAM boundaries and the overlapping of pulses.  As $\psi_0$ tends to moderately small or moderately large values, the resultant pulse-train consists essentially of double pulses (where the spacing between each pair of pulses is much larger than the spacing between each pulse in the pair).  It is well known that double pulses produce KAM boundaries (which are a classical effect), and this situation has previously been used to study the effects of quantum tunnelling through KAM boundaries \cite{Kendra,Greg}.  KAM boundaries should therefore explain the large scale decrease in energy in these regions, even for the quantum simulation.  This argument is supported by viewing the momentum distributions in these regions (as an example see figure \ref{fig:lowr130KAM}).  KAM boundary effects seem to become less important in the region $75^{\circ} \sim < \psi_0 \sim < 285^{\circ}$ where the energy seems to take on a more uniform value (ignoring resonance peaks).

The large peaks as $\psi_0 \rightarrow 0^{\circ},360^{\circ}$ are explained by the overlap of pulses.  At $\psi_0=0^{\circ},360^{\circ}$ the pulses are overlapping completely (periodic with period $30\mu$s) and there are only half as many resultant pulses as there would have been without overlaps.  Also $\kappa$ is twice as large as without overlaps.  Ignoring anomalous diffusion (and DL), the diffusion rate increases as $\kappa^2$ and the energy increases roughly linearly with kick number.  So with twice the kicking strength and half the number of pulses, the final energy obtained would be roughly two times as large as if there were no overlapping pulses.  This rough calculation is in approximate agreement with what is seen in our results.

It is interesting to note the positions of the resonance peaks (e.g. at $\psi_0=90^{\circ},120^{\circ},180^{\circ}$ in figure \ref{fig:highr130}).  They occur at rational fractions of $360^{\circ}$, i.e. when $\alpha_0$ (from Eq. (\ref{eq:ham1})) is a rational fraction.  These observations are suggestive of mode locking \cite{Lichtenberg}.  The larger peaks appear to follow the general pattern of the Farey tree \cite{Kim} in that the simpler the rational fraction, the larger the peak seems to be.

\begin{figure}[!h]
\centering
\includegraphics[height=5cm,width=8cm]{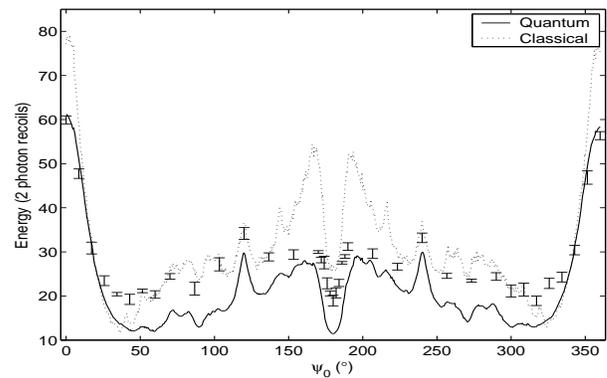}
\caption[Results for $r=1$ and $\kappa=10.1$]{Experimental and simulated (both quantum and classical) energies for $r=1$, $\kappa=10.1$, $N_{\rm{tot}}=30$ and $\eta=2.8\%$ as a function of $\psi_0$.  Errorbars are typically $\pm1.2$ energy units for the quantum simulation and $\pm1.7$ energy units for the classical simulation.}
\label{fig:r1low30}
\end{figure}

Figure \ref{fig:r1low30} shows the results  with $\kappa_1=\kappa_2=10.1$ and all other parameters the same as previously.  A similar large scale structure is obtained to that when $\kappa=17.7$ - large peaks at $\psi_0 \rightarrow 0^{\circ}, 360^{\circ}$, KAM boundary regions for moderately small and large $\psi_0$ and resonance peaks.  There are some differences however.  Some of the resonance peaks from figure \ref{fig:highr130} are now anti-resonance dips (e.g. $\psi_0=90^{\circ},180^{\circ}$) and there is a broad peak (with a dip in the middle) centred on $\psi_0=180^{\circ}$.

The distinction between quantum and classical behaviour is not so clear in the experimental data as for $\kappa=17.7$.  However, similarity with the quantum simulation is observed in the regions $\psi_0 \rightarrow 0^{\circ},360^{\circ}$ and $120^{\circ} < \psi_0 < 240^{\circ}$.  The qualitative differences between the quantum and classical simulations are expected to be largest in these regions, as the pulse train becomes more periodic for these initial phase offsets.

\begin{figure}[!h]
\centering
\includegraphics[height=5cm,width=8cm]{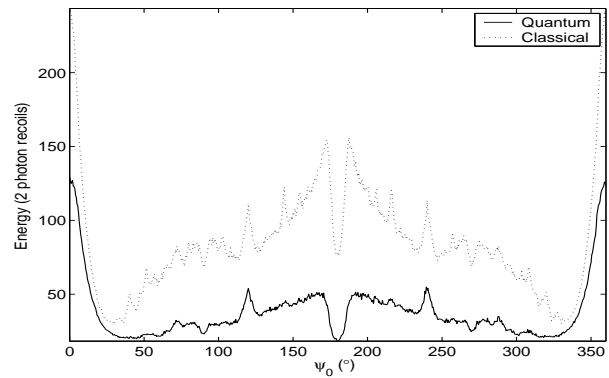}
\caption[Results for $r=1$ and $\kappa=10.1$]{Classical simulation for $\kappa=10.1$ and $\eta=2.8\%$ as a function of $\psi_0$ for $r=1$ and $100$ kicks.  Errorbars are typically $2.3$ energy units for the classical simulation and $\pm 3.7$ energy units for the quantum simulation.}
\label{fig:lowr1500}
\end{figure}

Quantum and classical simulations were run for $\kappa=10.1$ and $N_{\rm{tot}}=100$ (figure \ref{fig:lowr1500}) to see how these peaks/dips evolve in the quantum and classical cases.  In the classical simulation all peaks/dips are present and become sharper.  In the quantum simulation, however, only the peaks/dips that were largest for $N_{\rm{tot}}=30$ are present - the other resonances seem to have been washed out.  This suggests that the origin of these peaks/dips is classical, not quantum mechanical, and that quantum effects may destroy these resonances.  Another point to note is that the relative difference in energy between the quantum and classical simulations is much larger for $N_{\rm{tot}}=100$ than it was for $N_{\rm{tot}}=30$, suggesting the presence of effects of DL.  Note that atomic spontaneous emission leads to a finite late-time diffusion rate in the quantum case, but this rate is generally slower than the corresponding classical rate. Note also that, despite this breakdown of DL, the momentum distributions can remain essentially exponential in shape even after this number of kicks \cite{KendraNelson}.

In the classical simulation the peaks/dips are clearer after more kicks.  This is to be expected as the periodicity is resolved better after more kicks.  Furthermore, figure \ref{fig:lowr1500} shows that if $\alpha_0$ is written as an irreducible fraction, $\alpha_0=a/b$, the peaks always appear when $b$ is an odd number and the dips always appear when $b$ is an even number - regardless of whether $a$ is even or odd.  This observation held true even for a classical simulation for $N_{\rm{tot}}=500$, where peaks corresponding to less rational values of $\alpha_0$ were able to be resolved above the noise.

Poincar\'{e} sections were plotted to search for structures that may explain the resonances/anti-resonances.  Although there were structures present, clear correlation with the shape of the energy curve was not obvious.  Generally the island structures in phase space remained reasonably constant over large ranges of $\psi_0$ values.  An explanation for both the appearance and positions of these peaks and dips will require further investigation.

\subsubsection{Momentum Distributions for $r=1$}

\begin{figure}[!h]
\centering
\includegraphics[height=5.5cm]{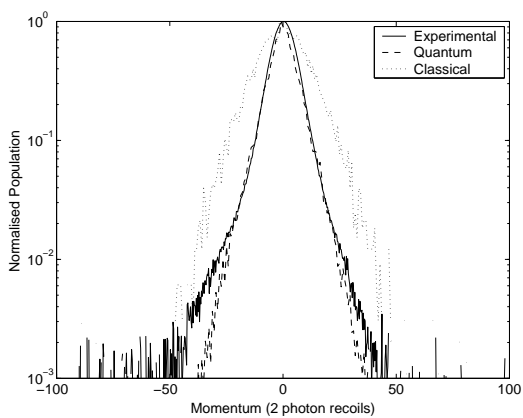}
\caption[Typical DL momentum distributions for $r=1$]{DL momentum distribution for $r=1$, $\kappa=10.1$, $N_{\rm{tot}}=30$, $\eta=2.8\%$ and $\psi_0=180^{\circ}$.  Note the experimental distribution has not been deconvolved with the initial position distribution, but this has only a slight effect on its shape.}
\label{fig:lowr130DL}
\end{figure}

As mentioned earlier, the difference in energy of the classical and quantum simulations suggest the presence of DL for both values of $\kappa$.  The key feature that distinguishes DL from other effects, however, is an exponentially localised momentum distribution.  Figure \ref{fig:lowr130DL} shows the momentum distributions obtained for $\kappa=10.1$ and $\psi_0=180^{\circ}$.  It still shows an essentially exponential momentum line shape, even for the level of spontaneous emission present, in good agreement with the quantum simulation.  A Gaussian distribution is seen for the classical simulation.  The distributions showing DL are most prominent around $\psi_0=120^{\circ},180^{\circ},240^{\circ}$, but are seen to a lesser degree for almost all values of $\psi_0$.

\begin{figure}[!h]
\centering
\includegraphics[height=5.5cm]{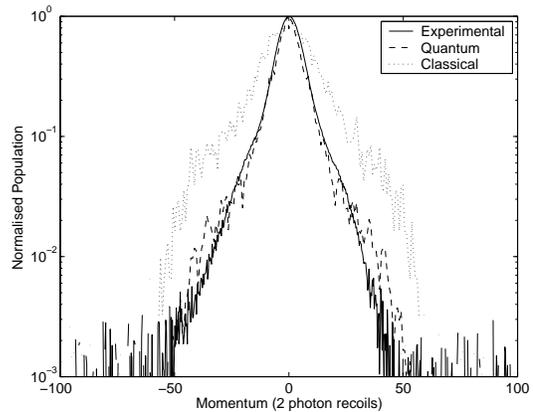}
\caption[Typical mix DL/KAM momentum distributions for $r=1$]{Mixed momentum distribution  for $r=1$, $\kappa=10.1$, $N_{\rm{tot}}=30$, $\eta=2.8\%$ and $\psi_0=170^{\circ}$.  Note the experimental distribution has not been deconvolved with the initial position distribution, but this has only a slight effect on its shape.}
\label{fig:lowr130DLAgain}
\end{figure}

In the region $110^{\circ}<\psi_0<250^{\circ}$ there is a mixture of DL and KAM boundary distributions (see figure \ref{fig:lowr130DLAgain} where an exponential distribution with shoulders is found, which is significantly narrower than the classical distribution and in agreement with the quantum distribution), except at $\psi_0=120^{\circ},180^{\circ},240^{\circ}$ where only DL is seen.  This suggests that KAM boundaries may even be having an effect on the dynamics in this region (it should be noted that the diffusion rate through a cantorus can be lower quantum mechanically than classically \cite{Kendra,Greg}).

Interestingly, at $\psi_0=120^{\circ},180^{\circ},240^{\circ}$ the resultant pulse-train contains (or very nearly contains) completely periodic pulses (i.e. having equal spacing).  The resultant pulse-train for $\psi_0=120^{\circ}$ may be viewed as being completely periodic with period $T=10\mu$s, but with every third pulse missing.  The resultant pulse-train for $\psi_0=180^{\circ}$ \emph{is} completely periodic with $T=15\mu$s.  $\psi_0=240^{\circ}$ is a similar case to $\psi_0=120^{\circ}$.

\begin{figure}[!h]
\centering
\includegraphics[height=10cm,width=8cm]{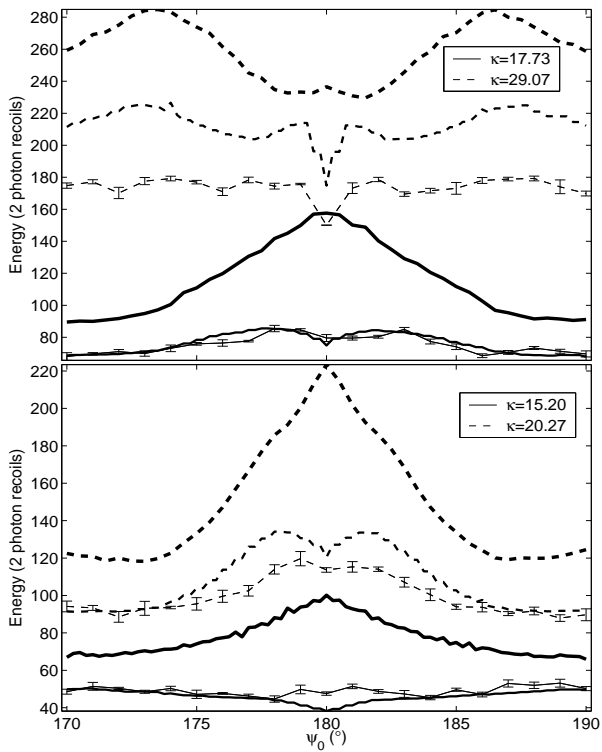}
\caption[Zoom-in results for $r=1$ and various $\kappa$]{Results for $r=1$ and $N_{\rm{tot}}=30$.  Classical simulations have the thickest lines, quantum simulations have thinner lines and experimental results have the thinnest lines with errorbars.}
\label{fig:zoom}
\end{figure}

The largest qualitative difference between the quantum and classical simulations is around $\psi_0=180^{\circ}$.  Although in figure \ref{fig:r1low30} the qualitative structure is very similar for the two simulations, in figure \ref{fig:highr130} there is a dip at $\psi_0=180^{\circ}$ for the quantum case and only a peak in the classical simulation.  This region was looked at in more detail for various kicking strengths.  The results are shown in figure \ref{fig:zoom} and indicate a reasonably complicated structure for different values of $\kappa$ for both quantum and classical dynamics.  The energy is lower in the quantum simulations (the peak has a dip in it), compared with the classical simulations, as $\psi_0 \rightarrow 180^{\circ}$.  This can only be explained by DL and indeed the momentum distribution exhibited an exponential lineshape.  Observing strong DL at this value of the initial phase is not surprising because, as mentioned earlier, the pulse-train is completely periodic.  Experimental results are always in qualitative agreement with the quantum simulation and show DL particularly around $\psi_0=180^{\circ}$.

\subsubsection{$r=1.4$}

\begin{figure}[!h]
\centering
\includegraphics[height=5cm,width=8cm]{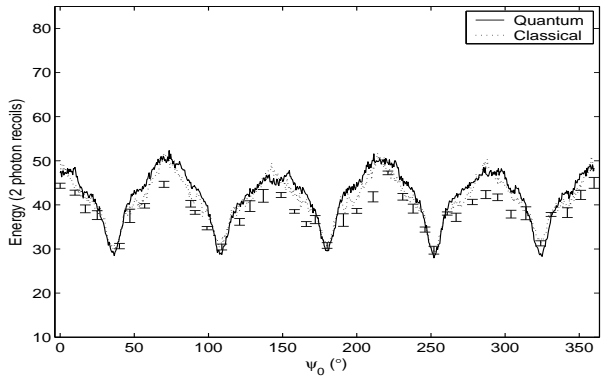}
\caption[Results for $r=1.4$ and $\kappa=10.1$]{Experimental energies along with quantum and classical simulations for $r=1.4$, $\kappa=10.1$, $N_{\rm{tot}}=30$ and $\eta=2.8\%$ as a function of $\psi_0$.  Errorbars are typically $\pm 1.7$ for the quantum simulation and $\pm 1.9$ for the classical simulation.}
\label{fig:lowr1.4}
\end{figure}

\begin{figure}[!h]
\centering
\includegraphics[height=5cm,width=8cm]{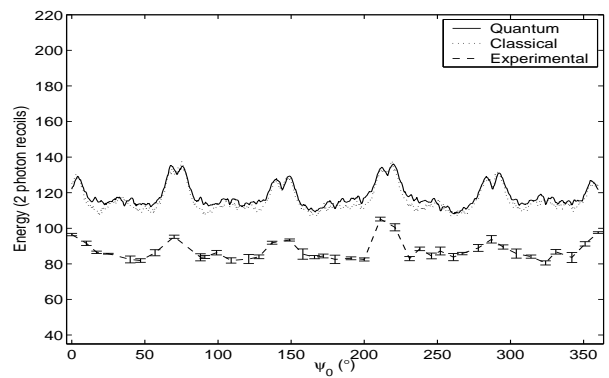}
\caption[Results for $r=1.4$ and $\kappa=17.7$]{Experimental energies along with quantum and classical simulations for $r=1.4$, $\kappa=17.7$, $N_{\rm{tot}}=30$ and $\eta=3\%$ as a function of $\psi_0$ for.  Errorbars are typically $\pm 1.7$ energy units for the quantum simulation and $\pm 2.6$ energy units for the classical simulation.  The experimental points have been joined to show the similarity in structure with the simulated results.}
\label{fig:highr1.4}
\end{figure}

Results for other two-frequency ratios were also obtained.  The experimental results for $r=1.4$ with $\kappa=10.1$ and $\kappa=17.7$, along with simulation results, are shown in figures \ref{fig:lowr1.4} and \ref{fig:highr1.4}, respectively.  Again the qualitative agreement between experiment and simulations is good.  There is definite structure present for both kicking strenghs which is shown in the experimental result as well as simulations.  However, for this value of the ratio there are no noticeable resonance peaks or anti resonance dips.  Also, there is no difference in energy between the classical and quantum simulations within numerical error.  It seems that for $r=1.4$, $30$ kicks is not enough to resolve the periodicity sufficiently for DL to have an effect.  This is backed up by the momentum distributions which do not show any evidence of DL.  They do, however, show that KAM boundaries are present.

These results show that even for this reasonably commensurate frequency, the periodicity is not resolved well enough on this time scale for the system to exhibit significant quantum effects.

\subsection{Incommensurate Frequencies}
\label{sec:incomfreq}

\begin{figure}[!h]
\centering
\includegraphics[height=5cm,width=8cm]{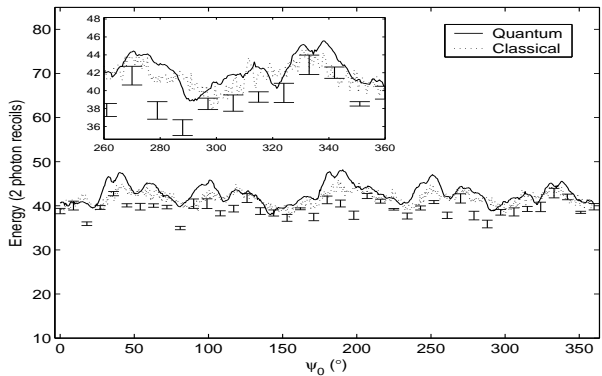}
\caption[Results for $r=\sqrt{2}$ and $\kappa=10.1$]{Experimental energies along with quantum and classical simulations for $r=\sqrt{2}$, $\kappa=10.1$, $N_{\rm{tot}}=30$ and $\eta=2.8\%$ as a function of $\psi_0$.  Errorbars are typically $\pm 1.7$ energy units for the quantum simulation and $\pm 2.1$ energy units for the classical simulation.}
\label{fig:lowroot230}
\end{figure}

\begin{figure}[!h]
\centering
\includegraphics[height=5cm,width=8cm]{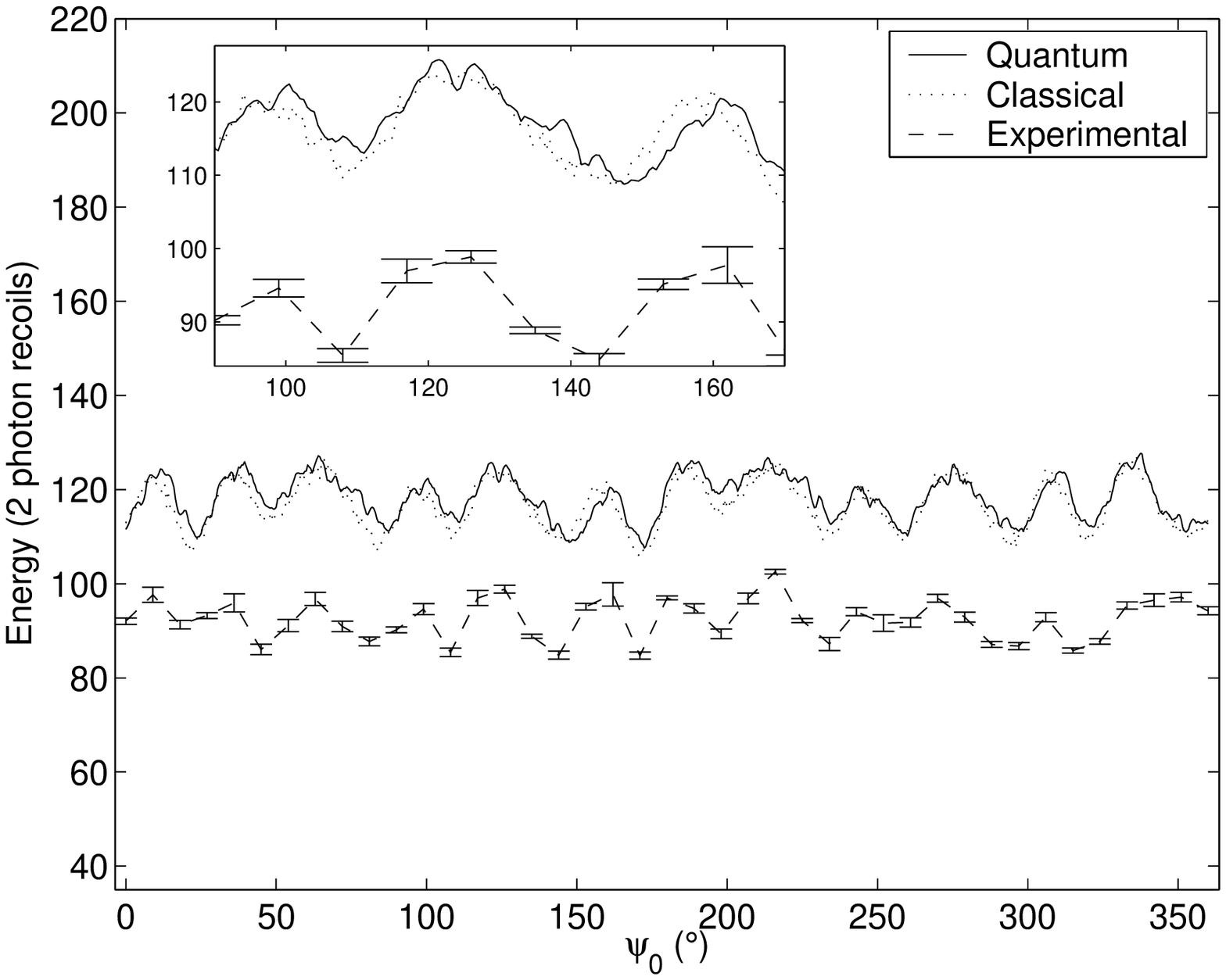}
\caption[Results for $r=\sqrt{2}$ and $\kappa=17.7$]{Experimental energies along with quantum and classical simulations for $r=\sqrt{2}$, $\kappa=17.7$, $N_{\rm{tot}}=30$ and $\eta=3\%$ as a function of $\psi_0$.  Errorbars are typically $\pm1.6$ energy units for the quantum simulation and $\pm 2.6$ energy units for the classical simulation.  The experimental points have been joined to show the similarity in structure with the simulated results.}
\label{fig:highroot230}
\end{figure}

If the ratio of pulsing periods is an irrational number, DL is not expected as the accumulated quantum phase from kick to kick will be approximately incoherent (for a sufficiently large number of kicks).  Thus the dynamics should be dominated by classical behaviour.

Figures \ref{fig:lowroot230} and \ref{fig:highroot230} show simulated and experimental results for $r=\sqrt{2}$ with $\kappa=10.1$ and $\kappa=17.7$, respectively, after $N_{\rm{tot}} = 30$ kicks.  Both quantum and classical simulations are shown.

As expected the energy shows a more uniform value compared to rational ratios.  Contrary to what may have been expected, the energy does show some dependence on $\psi_0$ and is not simply the flat line expected in the zero correlation limit.  There is definite structure present for both classical and quantum simulations, and the experimental results show this same structure.  The classical simulation is more noisy but despite this, the broader features of the quantum and classical simulations are identical within error.  This similarity could not be clearer than in the higher kicking strength case (see inset in figure \ref{fig:highroot230}).

The variations in energy, for both cases, seem to have some sort of systematic variation with initial phase.  When an FFT is performed the main spectral component is found to be $1 / \Delta \psi_0 \sim 1 / 70^{\circ}$ for $\kappa=10.1$ and $1 / \Delta \psi_0 \sim 1 / 30^{\circ}$ for $\kappa=17.7$.  The variations in energy, while small compared to the total energy (in the order of $10\%$), are nonetheless present.  The mean energy is around $41$ two photon recoils for $\kappa=10.1$.  This corresponds to the zero-correlation (quasilinear) limit, taking into account the spread in kicking strengths present in the real experiment due to the finite size of the beam and the cloud, as well as the magnetic sublevels of Caesium, and also the pulse-shape.

Figure \ref{fig:highroot230} also shows that the energy of the experimental data is lower than both the quantum and classical simulations.  This suggests that, at least for higher kicking strengths, we may be losing some of the atoms in the kicked cloud during the molasses freezing phase before imaging.  This energy offset can also be seen in other figures with higher kicking strengths.

The energy variations may be due to transient early time effects that have not yet been ``washed out'' after $30$ kicks.  Only a finite length pulse train can be applied to the rotors, and in theory an infinitely long train would be needed to realise an incommensurate frequency ratio properly.

This is related to the way in which an irrational number may be approximated arbitrarily well by an infinite sum of rational fractions.  The number of terms in the sum determines how well the irrational number is approximated.  The continued fraction representation of any number, $\omega$, between $0$ and $1$ may be written as:
\begin{equation}
\omega = \frac{1}{a_1+\frac{1}{a_2+\frac{1}{a_3+ ...}}}
\label{eq:irratapprox}
\end{equation}
An irrational number will have infinitely many non-zero integer terms, $a_i$.  The more terms that are included in the continued fraction, the better the approximation to the irrational number.  If an integer offset, $a_0$, is included, then any irrational number may be approximated by this method.

In terms of the aforementioned pulse train, due to the discretisation of time, more pulses is analogous to having more $a_i$ terms in the continued fraction representation.

\begin{figure}[!h]
\centering
\includegraphics[height=5cm,width=8cm]{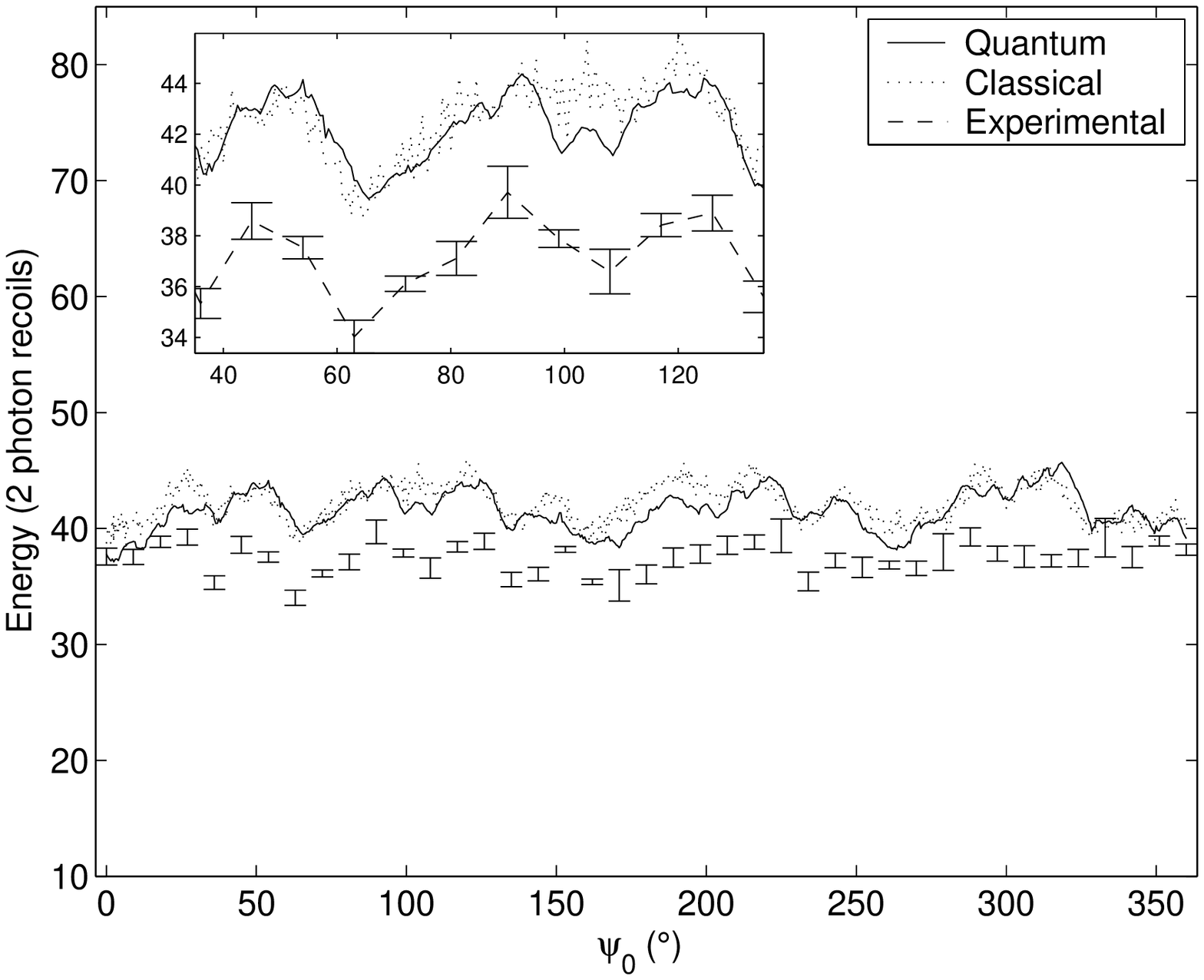}
\caption[Results for $r=\sqrt{3}$ and $\kappa=10.1$]{Experimental energies along with quantum and classical simulations for $r=\sqrt{3}$, $\kappa=10.1$, $N_{\rm{tot}}=30$ and $\eta=2.8\%$ as a function of $\psi_0$.  Errorbars are typically $\pm1.6$ energy units for the quantum simulation and $\pm 2.1$ energy units for the classical simulation.}
\label{fig:lowroot330}
\end{figure}

Figure \ref{fig:lowroot330} show results for $r=\sqrt{3}$ with all other parameters the same.  Slightly different structure is seen to that for $r=\sqrt{2}$, however the same trends are exhibited as for $r=\sqrt{2}$, i.e.  periodic variations on an approximately uniform background energy.

\begin{figure}[!h]
\centering
\includegraphics[height=5cm,width=8cm]{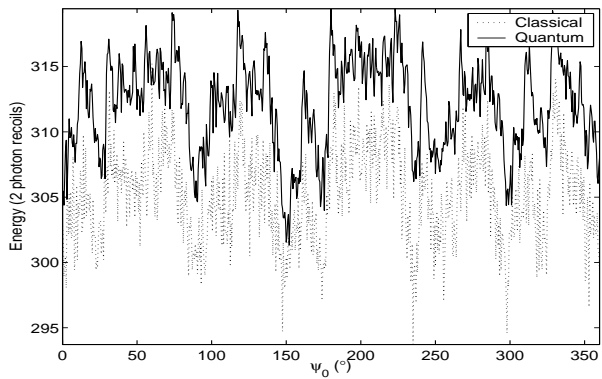}
\caption[Simulation results for $r=\sqrt{2}$, $\kappa=17.7$ and $90$ kicks]{Quantum and classical simulations of the energy for $N_{\rm{tot}}=90$, $r=\sqrt{2}$ and $\kappa=17.7$.  Errorbars are typically $\pm 5.2$ energy units for the quantum simulation and $\pm 6.0$ energy units for the classical simulation.}
\label{fig:highroot290}
\end{figure}

If the cause of the structure seen is the `small' number of kicks in the pulse-train, then having more kicks should cause the variations to be washed out.  It is not possible with the current experimental setup to test this theory, as not enough pulses can be stored in the memory of the PPG.  However it is possible to look at the effects found in simulation by increasing the kick number.  Figure \ref{fig:highroot290} shows the classical and quantum simulations for $\kappa=17.7$, $r=\sqrt{2}$ and $90$ kicks.  The structure is clearly different from that after $30$ kicks (figure \ref{fig:lowroot230}).  Four times the number of trajectories have been used here so as to keep the errors similar to the lower kicking number.  The absolute height of the variations is approximately the same as for the lower kicking number, however the variation relative to the total energy has decreased to around $3\%$ (compared to $10\%$ for the $30$ kick case).  There is less structure than in the low kick number case and it seems as though the energy is tending more towards a more uniform value as the kick number is increased.  As more kicks are added to the pulse-train, it is likely these transient effects would become less and less noticeable.

Weak KAM boundary distributions are seen for practically all irrational values of $r$ and all $\psi_0$  in the simulations (however the boundaries are not nearly as strong as in the cases with rational $r$) and no DL is seen in the quantum simulation or the experimental results.  However, the KAM boundaries were too weak to be observed in the experimental momentum distributions.  It is not clear that the KAM boundaries are responsible for the observed structure, however it is clear that the dynamics for irrational ratios is again governed by classical effects.

\subsection{Energy Growth as a Function of the Ratio of the Driving Frequencies}
\label{sec:ratio}

\begin{figure}
\includegraphics[height=10cm,width=8cm]{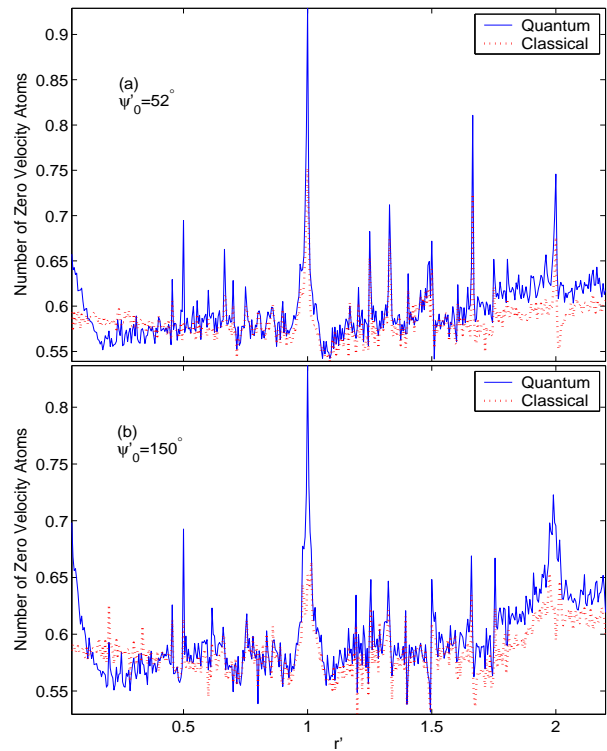}
\caption{Simulations of the system described in \cite{Frenchgroup} with $N_{\rm{tot}}=50$ and $\kappa=10$.  The number of zero velocity atoms has been calculated by integrating the normalised final momentum distribution over $-\epsilon$ to $\epsilon$ for some fixed value of $\epsilon$.}
\label{fig:french}
\end{figure}

As mentioned earlier, when the ratio of the driving frequencies was highly rational, Ringot et al. \cite{Frenchgroup} found an increase in the number of zero velocity atoms compared to when it was highly irrational.  This behaviour was attributed solely to DL.  Although the parameters of ratio and initial phase have been defined differently here than in \cite{Frenchgroup}, they are simply related by $r'=1/r$ and $\psi'_0 = \psi_0 / r$.  Using these conversions, it can be seen from figure \ref{fig:r1low30} that for $r=r'=1$, $\kappa \approx 10$ and $\psi'_0=\psi_0=52^{\circ}$ (the specific phase used in \cite{Frenchgroup}), KAM boundaries play a large role in the dynamics and in fact also halt the energy growth.  Hence the number of zero velocity atoms would be larger than if KAM boundaries were not present.  The case where KAM boundaries are weak can be seen in figure \ref{fig:lowroot230} ($r'=1/\sqrt{2} \approx 0.707$) with $\psi_0=52^{\circ} \times \sqrt{2} \approx 73.5^{\circ}$.  Here the energy is larger than the previous case and hence fewer zero velocity atoms are expected.  This is true for both the quantum and classical cases and hence both cases should show an increased number of zero velocity atoms for $r'=1$ compared to $r'=\sqrt{2}$.

To verify this we also ran quantum and classical simulations for parameters equivalent to those stated in \cite{Frenchgroup}.  The results are shown in figure \ref{fig:french}a.  In our simulations both the quantum and classical cases display peaks in the number of zero velocity atoms for rational values of $r'$.  Also, we examined the momentum distributions to determine whether DL or KAM boundaries were responsible for these peaks.

In contrast to the explanation given in \cite{Frenchgroup}, by examination of the simulated momentum distributions, we find that KAM boundaries, and not DL, are primarily responsible for the increase in the number of zero velocity atoms.  Only a small amount of DL is observed for the most rational ratios and even then the most dominant effect is that of KAM boundaries.

The fact that the effects of DL seem to be easily destroyed by changing the initial phase offset suggests that the results obtained in \cite{Frenchgroup} were to some degree due to the particular choice of initial phase and may not be true in general.  To check for this we also ran simulations with $\psi_0'=150^{\circ}$.  The results are shown in figure \ref{fig:french}b.  There is a difference between this case and the $\psi_0'=52^{\circ}$ case in terms of structure - particularly around $r'=2$.  Peaks still occur at rational values of the ratio in both the classical and quantum cases.  The peaks in the classical and quantum cases are approximately the same height, except for the peak at $r'=1$, which is noticeably lower in the classical case.  The quantum momentum distribution at this value of $r'$ distinctly shows DL, as well as KAM boundaries.

\section{Conclusion}
\label{sec:Conclusion}

To summarise, for rational ratios, those values of the initial phase that give a more periodic resultant pulse-train do show DL on the timescale investigated.  However, the values of the initial phase that resulted in less periodic pulse-trains do not show DL on the timescale investigated, even for the most rational ratios.

The initial phase offset of the resulting pulse-train has a large effect on the dynamics in both the quantum and classical cases for rational ratios, and this effect is enhanced on longer timescales.  The only effect present for irrational ratios is a weaker transient structure that has a diminished effect on longer timescales.
 
No DL is seen when the ratio, $r$, is irrational for any value of the initial phase offset.  That is, quasi-periodic driving destroys quantum kick to kick correlations on the timescale investigated.  

Most of the observed effects have been found to be classical, not quantum mechanical, in origin and are not due to dynamical localisation.  A dominant factor in the dynamics is KAM boundaries.

T.M. is grateful to the Computer Science department of the University of Auckland for the use of their $360$ node computer cluster.  This work was supported by the Royal Society of New Zealand Marsden Fund,
 grant UOA016.

\end{document}